\documentclass[12pt]{iopart}
\usepackage{graphicx}
\usepackage{dcolumn}
\usepackage{bm}
\usepackage{slashbox}
\usepackage{iopams}  
\begin{document}
\title[Nonadditivity in binary charged colloidal
  suspensions]{Nonadditivity in the effective interactions of binary
  charged colloidal suspensions} 

\author{E Allahyarov$^{1,2}$, H L{\"o}wen$^1$}
\address{ $^1$ Institut f\"ur Theoretische Physik II: Weiche Materie,
Heinrich-Heine-Universit\"at D\"{u}sseldorf, 
Universit{\"a}tsstra{\ss}e 1, D-40225 D\"{u}sseldorf, Germany}
\address{$^2$ Department of Physics, Case Western Reserve University, Cleveland,
Ohio 44106, USA,  and Joint Institute for High Temperatures, Russian Academy of Sciences, Moscow, Russia}
\ead{elshad.allakhyarov@case.edu}
\begin{abstract}
Based on primitive model computer simulations with explicit microions, we calculate the
effective interactions in a binary mixture of charged colloids
with species $A$ and $B$ for different size and charge ratios.
An optimal pairwise interaction is obtained by fitting the many-body effective forces. This 
interaction is close to a Yukawa (or Derjaguin-Landau-Verwey-Overbeek(DLVO)) pair potential  but
the $AB$ cross-interaction is different from  the geometric mean of the two direct $AA$ and $BB$
interactions. As a function of charge asymmetry, the  corresponding
nonadditivity parameter is first positive, then getting significantly negative 
and is getting then positive again. 
We finally show that an inclusion of nonadditivity
within an optimal effective  Yukawa model gives better predictions 
for the fluid pair structure than DLVO-theory.
\end{abstract}
\submitto{\JPCM, special issue}
\maketitle

\section{Introduction}
Phase diagrams and structural correlations in binary mixtures are much richer
than those of their one-component counterparts \cite{Tammann} 
since there are
additional thermodynamic degrees of freedom. Understanding the phase behaviour 
from first principles \cite{Hafner,book} requires the knowledge
of the effective interaction forces between the different species which
is - in general - a many-body force. Even if this
interaction is pairwise additive, the full calculation of structure and phase behaviour
has only been done for selected cases. Among those are
 hard spheres \cite{Pronk,Xu,Eldridge1,Bartlett1}, oppositely charged colloids
\cite{Leunissen_Nature_2005,Hynnien_PRL_2006},
two-dimensional dipolar mixtures \cite{Assoud,Likos_2007} and two-dimensional Yukawa mixtures 
\cite{Assoud_2}.

In this paper we consider a three-dimensional binary colloidal suspension
of two species $A$ and $B$ of charged spheres ("macroions") with different charges 
($Z_A e$ and $Z_B e$, $e$ denoting the electron charge) 
and diameters ($\sigma_A$ and $\sigma_B$) \cite{review,Levin,Messina}.
The traditional Derjaguin-Landau-Verwey-Overbeek (DLVO) theory describes  the 
interaction between the two species as an effective pairwise Yukawa potential 
\cite{Klein1,Klein2}
\begin{equation}
V(r) = \frac{Z_i e}{1+ \sigma_i \kappa_D/2} \frac{Z_j e}{1+ \sigma_j
  \kappa_D/2} \frac{\exp ( -\kappa_D (r - (\sigma_i + \sigma_j)/2) )} {\epsilon r} 
\label{eq1}
\end{equation}
where $(ij)= (AA), (AB), (BB)$, $\epsilon$ denotes the dielectric constant of the solvent 
and $\kappa_D$ is the Debye-H\"uckel screening parameter. The latter is given as
\begin{equation}
{\kappa_D}^2 = 4\pi (\sum_j z_j^2\rho_j) /\epsilon k_BT  
\label{eq2}
\end{equation}
where the sum runs over all microions
with their charges $z_j$ and partial number densities $\rho_j$.
The DLVO theory is a linearized theory and therefore neglects nonlinear
 screening effects \cite{LMH1,LMH2} which give rise to effective many-body 
forces \cite{triplet,Wu,Reinke,Russ,Kreer}. Nonlinear effects can at
least partially be accounted for  
by charge renormalization which is conveniently calculated in a spherical 
Poisson-Boltzmann cell model \cite{Trizac}. The
 cell approach was recently generalized towards binary mixtures  by  Torres, T\'ellez and  
van Roij \cite{Torres}. In the latter approach, it was shown that charge renormalization is different
for the different species such that the ratio of effective charges is different from that
of the bare charge. However, the cross-interaction was not addressed in this study.

The importance of the cross-interaction between $A$ and $B$ relative to the direct part
$AA$ and $BB$ determines the so-called non-additivity parameter $\Delta$ of the mixture which 
is crucial for the topology of phase diagrams. Binary hard sphere systems
have been studied as a prototype for any non-additive mixtures \cite{nonadd_1,nonadd_2,nonadd_3}.
In general, a positive non-additivity is realized if the cross-interaction is more repulsive than 
the mean of the two direct interactions. For high positive 
non-additivity ($\Delta >0$), macrophase separation
into an $A$-rich and $B$-rich phase is observed, i.e. the system minimizes the interface where the
$AB$ cross-interactions plays a dominant role. The other case of negative nonadditivity ($\Delta <0$)
implies a weaker cross-interaction in terms of the bare ones such that the system tends to mix and
to exhibit micro-phase-separation \cite{Hoffmann2}. 

For pairwise Yukawa interactions $V_{ij}(r) ={Z_{ij}^*}^2\exp(-\kappa_D
r)/r$ ($(i,j) = (AA), (AB), (BB))$,   
a dimensionless nonadditivity
parameter $\Delta$ can be quantified 
by invoking the deviation of an ideal Berthelot mixing rule
\cite{Hopkins1,Hopkins2} via the relation 
\begin{equation}
 1 + \Delta  = {Z_{12}^*}^2 / {Z_{11}^*}{Z_{22}^*}  
\label{eq3}
\end{equation}
 For charged suspensions, classical DLVO theory (see eqn. (1)) implies a vanishing $\Delta$
since the effective charges are the same in all interactions.
There are other realizations of a binary Yukawa system in  dusty plasmas
\cite{Vaulina,Kalman_PRL_2004,Sutterlin}
and  metallic mixtures \cite{Horbach1} or amorphous silica \cite{Horbach2}.
In fact, the binary Yukawa model has been widely used and employed to investigate
 effective interactions \cite{Louis},
fluid-fluid phase separation \cite{phase1,phase2,phase3},
 vitrification \cite{polydis,dynamics1,dynamics2,Chavez,Medina}  and
 transport properties \cite{Salin}.  
In most of studies of binary Yukawa systems, 
the non-additivity parameter is set to zero, except for Refs.\ \cite{Hopkins1,Hopkins2}
where the effect of positive nonadditivity $\Delta$ on fluid-fluid phase separation is considered.
In the context of dusty plasmas, there is another recent study showing
 that the non-additivity parameter $\Delta$ is positive in general \cite{Ivlev}.
This leads to macrophase separation in binary dusty plasmas, as observed 
in experiments \cite{Sutterlin,Ivlev}.
The physical origin of the interaction in dusty plasmas, however,
 is different from that relevant for 
charged colloidal suspensions. While for the former the ion are described by a Gurevich distribution,
a Boltzmann distribution is appropriate for the latter. 

In this paper,
we focus on the nonadditivity of the cross-interaction for charged colloidal suspensions.
Using computer simulations with explicit microions \cite{Amico1,Amico2,Trigger,Damico}, we calculate 
the effective interactions 
in a charged binary mixture and find that the sign of the  non-additivity depends 
on the parameters, in particular on the charge asymmetry.
The nonadditivity parameter is calculated first by using simulations
of three pairs of macroions, namely 
$AA$, $AB$, and $BB$ in a periodically repeated simulation box at
fixed screening. We also consider  
larger systems with 24 macroions at different compositions in order to
check the effect on many-body forces 
on $\Delta$. In the latter case we fit the many-body forces by  effective pairwise forces
$- dV_{ij}(r)/dr$ and extract $\Delta$ from the optimal fit  \cite{Kramposthuber}.
We confirm that $\Delta$ is unchanged. Our main findings are i)  that $\Delta$ is typically 
large and cannot be neglected and ii) that the sign of $\Delta$ depends on the parameters 
as e.g. charge asymmetry. If a binary charged colloidal mixture is described by an effective
Yukawa model, $\Delta$ needs to be incorporated into the description. For instance when 
compared to DLVO theory a much better description of the fluid pair structure is achieved
within an effective Yukawa model and non-vanishing $\Delta$.

The paper is organized as follows: In Sec. II  we describe the model and the simulation method
and apply it to the case of two macroions. Many-body simulations with 24 macroions 
are discussed in Sec. III. Then we present data for a large system
with effective pairwise Yukawa forces 
in order to see the effect of non-vanishing nonadditivity in Sec. IV. Finally we conclude in Sec. V.

\section{Simulations with two macroions}
We model all ions as uniformly charged hard
spheres such that they are interacting via excluded volume and Coulomb
forces which are reduced by the dielectric constant $\epsilon$ of the
solvent.  The two species of charged colloids have a mesoscopic hardcore diameter
$\sigma_A$ (resp.\ $\sigma_B$) and a total charge $Z_A e$ (resp.\ $Z_B e$)
while all microions are monovalent
with a charge $e$ ($e$ denoting the electron charge) and a microscopic
hardcore diameter $\sigma_c$.  For finite salt concentrations,
there are both counter- and coions in the solution and the 
microscopic core of oppositely charged microions is
needed to prevent the system from the Coulomb collapse.  
The averaged concentration of added salt is denoted
with $n_s$. The salt is always monovalent. The system is kept at room temperature $T$ such that the
Bjerrum length for the microions is $\lambda_B = e^2/\epsilon
k_BT=7.8$\AA \,  
with $\epsilon =80$ the dielectric constant of water at room temperature.

A cubic simulation box of edge length $L$ 
with periodic boundary conditions is used containing two macroions
and the following three cases are studied separatedly: i) two $A$ macroions,
ii) two $B$ macroions, and iii) one $A$ and one $B$ macroion. The two macroions
are placed along the room diagonal of the simulation box and possess a fixed central distance $r$.
At fixed macroion positions, the microions are moved by constant temperature molecular dynamics
and the averaged force $F$ acting on the two macroions is
calculated. The latter fulfills Newton's third law. 
For more technical details we refer to the Refs.\ \cite{Amico1,Amico2,wedge,protein1,protein2,Zacca}.
Then the distance $r$ is varied and force-distance curve $F(r)$ is gained.

Inspired by the DLVO expression (1), we anticipate that the screening length will not differ much 
in the three cases i), ii), iii) and that it will be comparable to the Debye-H\"uckel
expression. This assumption will be tested and justified 
later by a many macroion simulation reported in Section III.
Therefore we adjust the box length in the three cases in order to reproduce the same
Debye-H\"uckel screening length (2).

\begin{figure} [!ht]
\hspace{-1.cm}
\includegraphics*[width=0.95\textwidth]{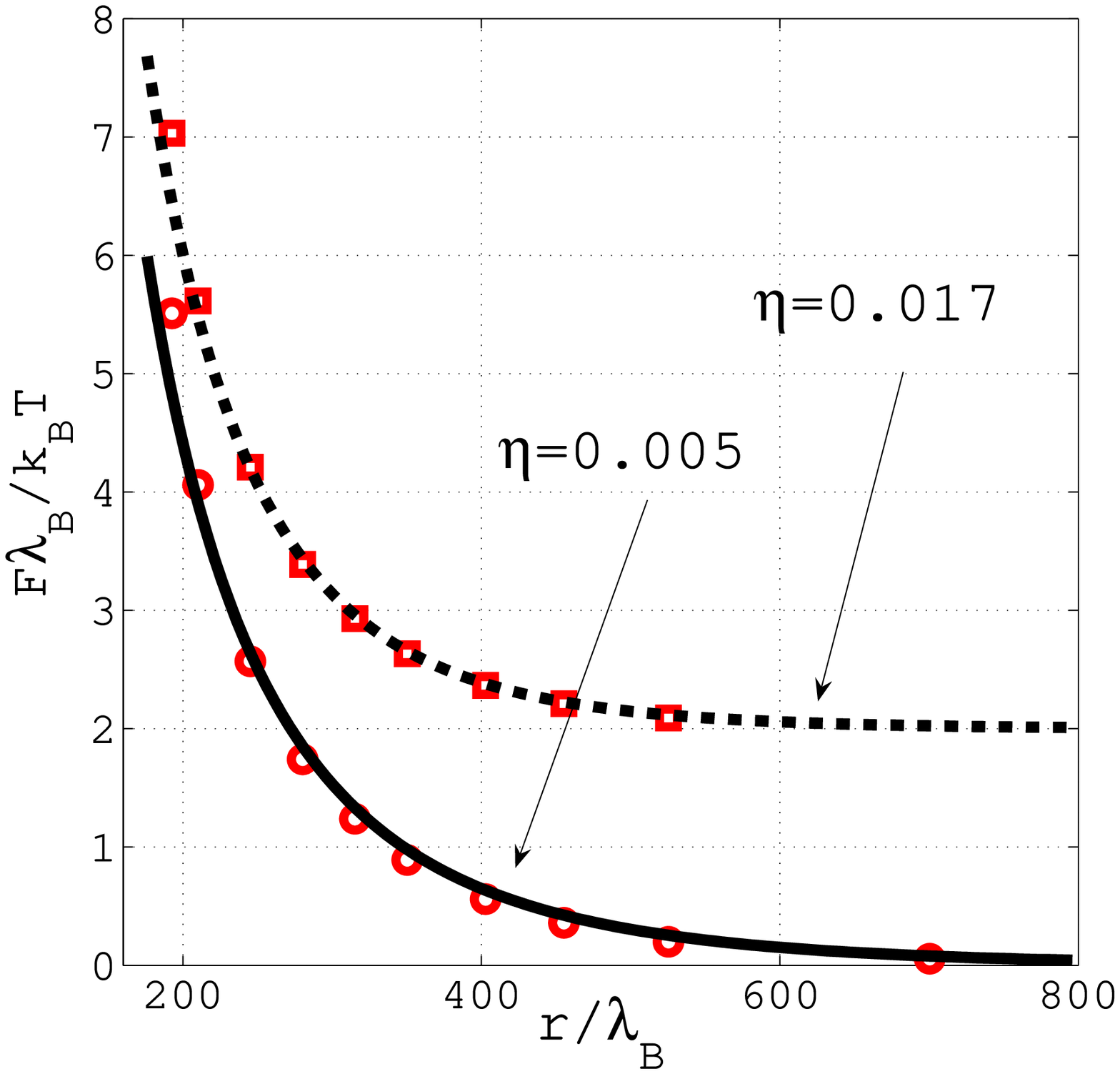}
 \caption{
(Color online) Dimensionless interaction force $F\lambda_B/(k_BT)$ for
   case i) ($AA$ macroion pair)  as
   a function of  dimensionless separation 
   distance $r/\lambda_B$. Symbols denote the simulation data, the
   full curves are the Yukawa fit for two macroion packing  fractions
      $\eta=0.005$ (soilid line) and $\eta=0.017$ (dashed line, shifted upward). The fit data are
   given in the text. 
\label{fig-1}
}
\end{figure}

\begin{figure} [!ht]
\hspace{-1.cm}
\includegraphics*[width=0.95\textwidth]{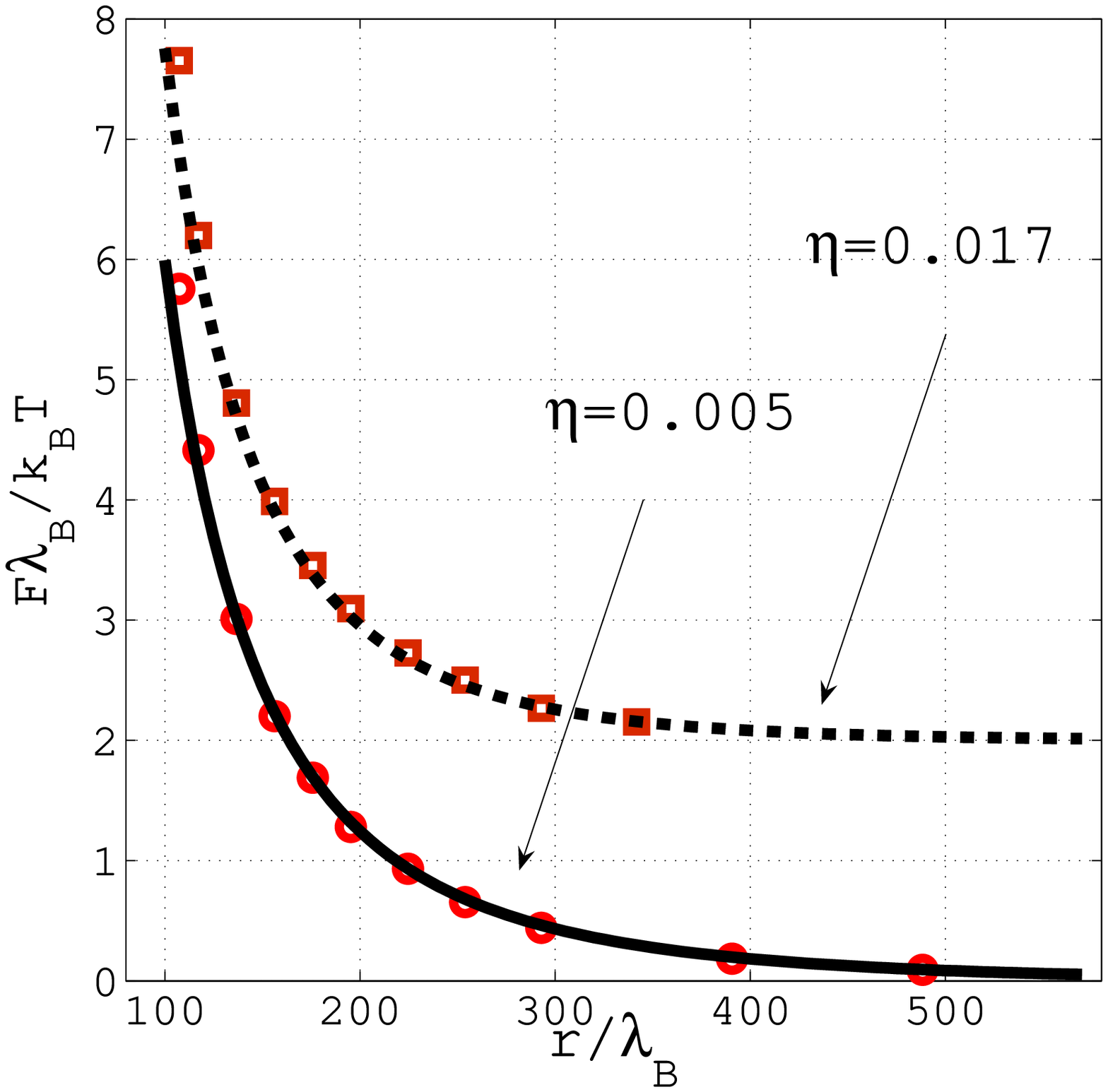}
 \caption{
(Color online) Dimensionless interaction force $F\lambda_B/(k_BT)$ for
   case ii) ($BB$ macroion pair)  as
   a function of a dimensionless separation 
   distance $r/\lambda_B$. Symbols denote the simulation data, the
   full curves are the Yukawa fit for two macroion packing  fractions
      $\eta=0.005$ (solid line) and $\eta=0.017$ (dashed line, shifted upward). The fit data are
   given in the text. 
\label{fig-2}
}
\end{figure}

\begin{figure} [!ht]
\hspace{-1.cm}
\includegraphics*[width=0.95\textwidth]{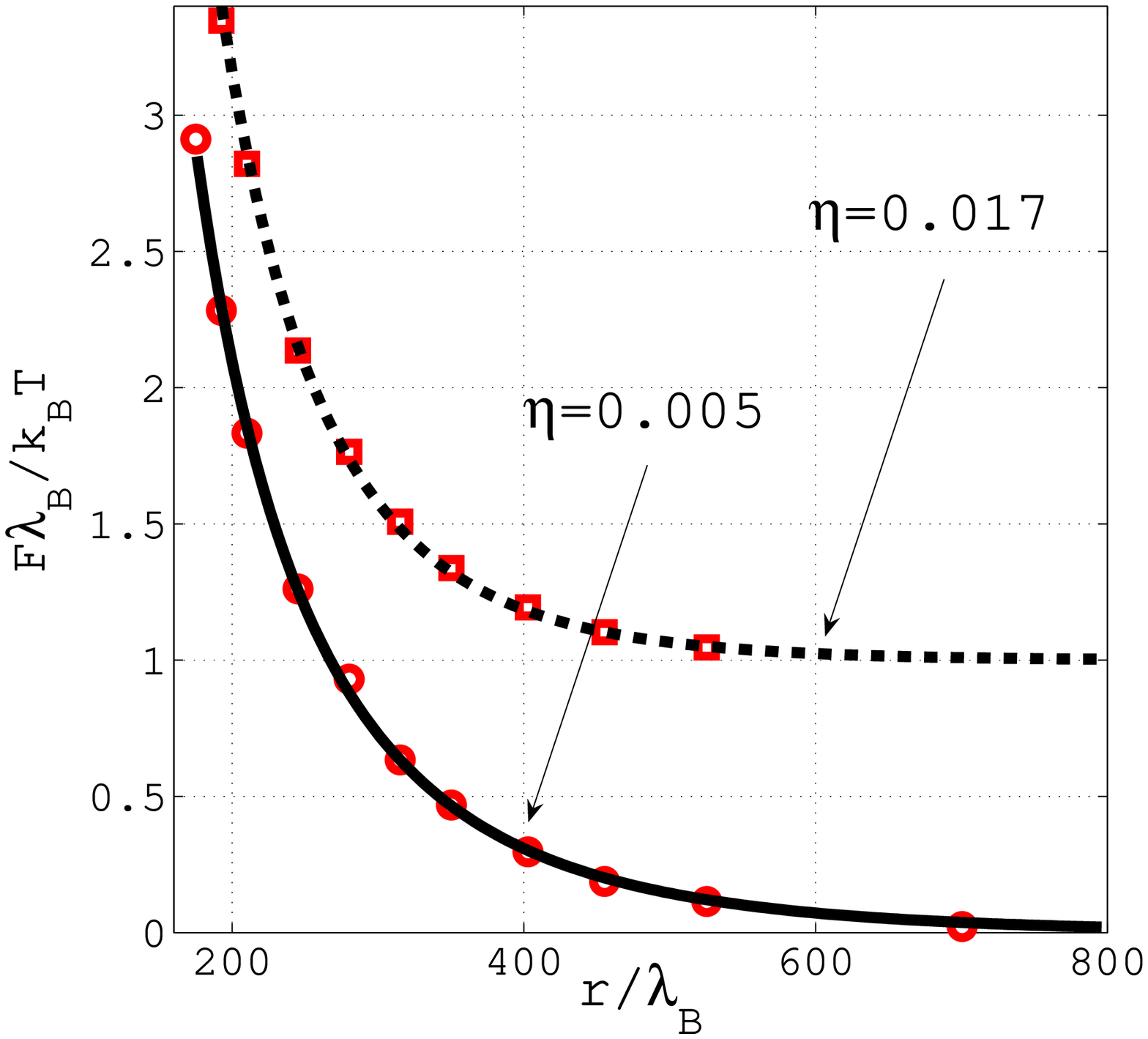}
 \caption{
(Color online) Dimensionless interaction force $F\lambda_B/(k_BT)$ for
   case iii) (pair of $A$ and $B$ macroion)   as  a function of a dimensionless separation 
   distance $r/\lambda_B$. Symbols denote the simulation data, the
   full curves are the Yukawa fit for two macroion packing  fractions
      $\eta=0.005$ (solid line) and $\eta=0.017$ (dashed line, shifted upward). The fit data are
   given in the text.
\label{fig-3}
}
\end{figure}

Results for the distance-resolved forces $F(r)$ are shown in Figures~\ref{fig-1}--\ref{fig-3} for 
the three cases i), ii), and iii) for extremely dilute macroion
suspension with a packing fraction $\eta=0.005$. Such dilute case is
chosen to diminish the boundary effects of the simulation box on the
interaction forces. The parameters used are $\sigma_A$=1220 \AA, $\sigma_B$=680\AA,
$Z_A=580$, and $Z_B=330$. The prescribed screening is $\kappa_D \sigma_A$= 0.8
corresponding to box lengths of $L=6.24\sigma_A$ (for i)), $L=5.72\sigma_A$ (for ii)), and
$L=6.0\sigma_A$ (for iii)). The obtained data for the distance-resolved forces $F(r)$ were
fitted with the Yukawa expression ${Z_{ij}^*}^2 e^2/(\epsilon r) (1/r + \kappa) \exp (-\kappa r)$
with $(ij)= (AA), (AB), (BB)$. The screening parameter $\kappa$ and
the three effective charge numbers ${Z_{AA}^*}$, ${Z_{BB}^*}$, and ${Z_{AB}^*}$ 
are used as fit parameters. We obtain $\kappa \sigma_A = 0.81$ (very close to its Debye-H\"uckel expression)
and effective charge numbers of  ${Z_{AA}^*}$=470, ${Z_{BB}^*}$=260, and ${Z_{AB}^*}$=330
such that the nonadditivity parameter is negative:
\begin{equation}
\Delta =  \left({Z_{AB}^*} \right)^2 / \left( {Z_{AA}^*} {Z_{BB}^*}  \right) - 1 = 
-0.11 
\label{nonadd}
\end{equation} 

In a similar manner we calculate the interaction forces and Yukawa
fitting parameters for another packing fraction $\eta=0.017$ at
which the images of macroions in neighboring cells start to affect the
long-range macroion-macroion interactions. We obtain $\kappa_D \sigma_A = 1.15$ 
and effective charge numbers of  ${Z_{AA}^*}$=505, ${Z_{BB}^*}$=265, and ${Z_{AB}^*}$=342
such that the nonadditivity parameter is $\Delta$=-0.13.  

The results for extreme dilute $\eta=0.005$ and dilute $\eta=0.017$
cases shown in Figures~\ref{fig-1}--\ref{fig-3} reveal that first of all, the
Yukawa expression for the forces is an excellent fit over the relevant distance range explored.
Moreover, while the  screening parameter is very close to its Debye-H\"uckel expression,
the nonadditivity of 11--13 percent is significant.

\begin{figure} [!ht]
\hspace{-1.cm}
\includegraphics*[width=0.95\textwidth]{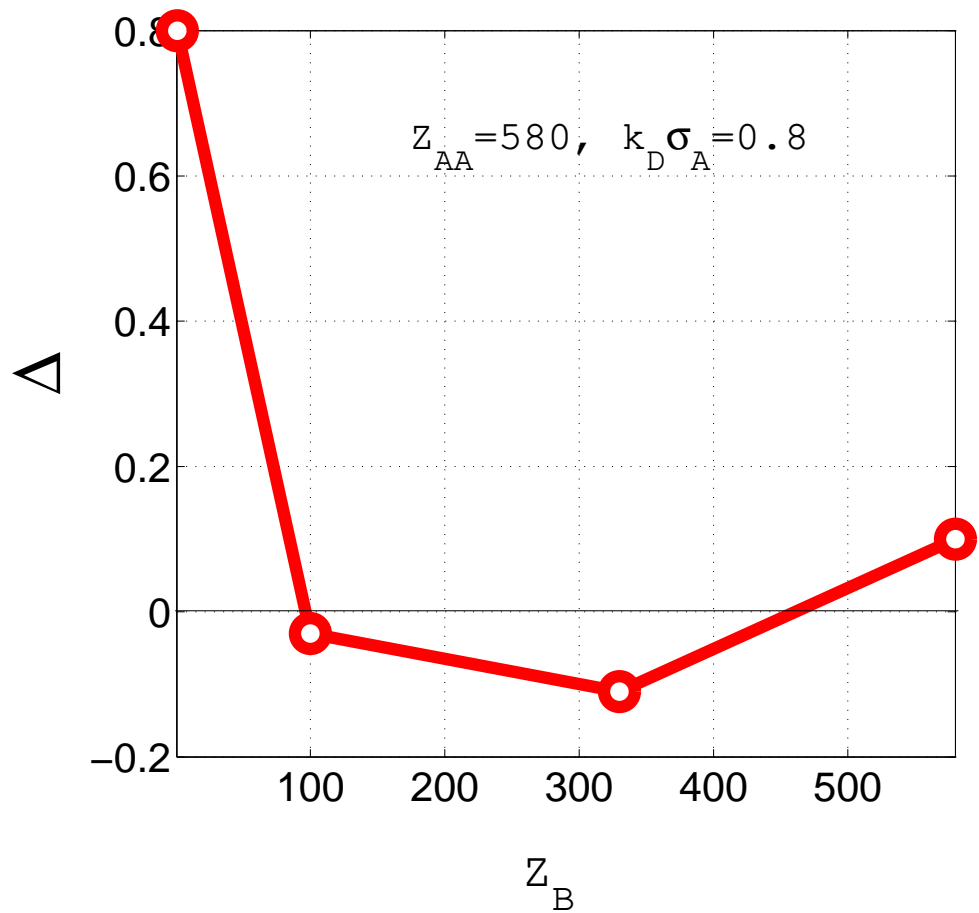}
 \caption{
(Color online)  Nonadditivity parameter $\Delta$ for different charges $Z_B$
   at a fixed charge $Z_A=580$. 
\label{fig-4}
}
\end{figure}

Next we explore the dependence of $\Delta$ on the charge asymmetry $\alpha=Z_A/Z_B$ 
by changing it in the range from 0 to
1 while keeping the
size asymmetry $\sigma_A/\sigma_B$=1.8 unchanged. In detail, we consider the
$B$-charges $Z_B$=0, 100, 330, 580 with fixed  $Z_A=580$ and $\kappa_D \sigma_A$=0.8.
Using the same simulation technique and fitting procedure, we extract 
the nonadditivity parameter $\Delta$ according to
Eq.(\ref{nonadd}). The results are presented in Figure~\ref{fig-4}.
The nonadditivity parameter $\Delta$ shows a clear non-monotonicity as a function of $Z_B$
starting from positive values and turning to negative ones and back to 
positive ones as $Z_B$ is increasing.
For $Z_B=0$, the $BB$ interaction is small,
but the $AB$ interaction has still a repulsive contribution from entropic contact force
\cite{Trigger} which drives $\Delta$ altogether towards a positive value.
The other cases are less intuitive and we do not have a simple argument for the sign of 
$\Delta$.

We  have finally considered also the case of size-symmetric ($\sigma_A =
\sigma_B$) but   charge-asymmetric $Z_A/Z_B$=1.76 macroions
at a fixed screening length of $\kappa_D\sigma_A=0.8$. Here,
the nonadditivity parameter $\Delta$  was found to be
-0.01, much smaller than for the corresponding size-asymmetric case 
with $\sigma_A/\sigma_B=1.8$. Hence  size-asymmetry appears to be the more crucial
input for the nonadditivity.

\section{Simulations with many macroions}
\begin{figure} [!ht]
\hspace{-1.cm}
\includegraphics*[width=0.95\textwidth]{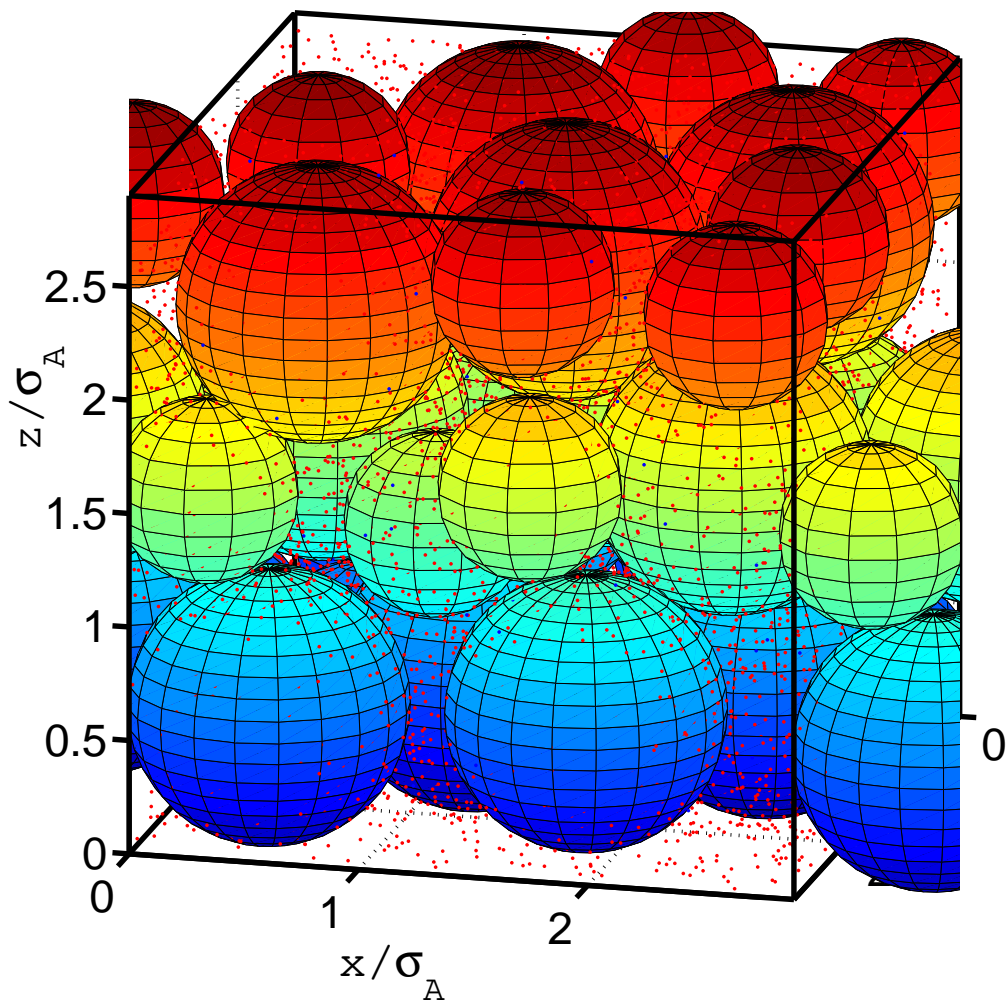}
 \caption{
(Color online) Full system  snapshot picture for 24 macroions  with
   equimolar composition $X=1/2$. The system size  is  2.9 $\sigma_A$ at a total packing fraction of
   $\eta$=0.3. The positively charged  counter- and salt ions  are shown as red 
   dots, while the negatively salt ions are shown as blue dots.  A vertical color
   gradient has been used for macroion positions along the $z$-axis. The parameters are:
   $Z_A$=580, $Z_B$=330.
\label{fig-7}
}
\end{figure}

Let us now turn to a many body simulation of the primitive model in a cubic cell
containing altogether $N=N_A + N_B = 24$ macroions with different compositions
$X=N_B/(N_A + N_B)$. The simulation box
contains also  $N_c= N_A Z_A + N_B Z_B$ oppositely charged counterions,
and $N_{s}$ salt ion pairs. The other parameters are as before if not otherwise stated.
  Six different macroion packing
fractions were considered: $\eta$= 0.017, 0.034, 0.12, 0.16, 0.23,
0.3.  For all simulations the salt concentration was kept constant at 
$n_{s}$=4$\times10^{-6}$ mol/l.  
Now both the microions and the macroions 
are moved by constant temperature molecular dynamics. A simulation snapshot
is shown in  Figure~\ref{fig-7}.
After equilibration, we stored 200 statistically independent macroion
configurations for each run.
For each stored configuration, we averaged the total forces $\vec F_i$ ($i$=1, $N$)
acting on the $i$th macroion 
over the microionic degrees of freedom. These forces are clearly many-body forces, in general.
Following the idea of Ref.\ \cite{Kramposthuber}, we assign an optimal effective pair interaction
by fitting all forces $\vec F_i$ in all stored configurations by the same  pairwise
Yukawa interaction. As in Section II, the four fitting parameters are 
$\kappa_D$ and
the three effective charge numbers ${Z_{AA}^*}$, ${Z_{BB}^*}$, and ${Z_{AB}^*}$ 
which determine the nonadditivity $\Delta$ directly. 

\begin{figure} [!ht]
\hspace{-1.cm}
\includegraphics*[width=0.95\textwidth]{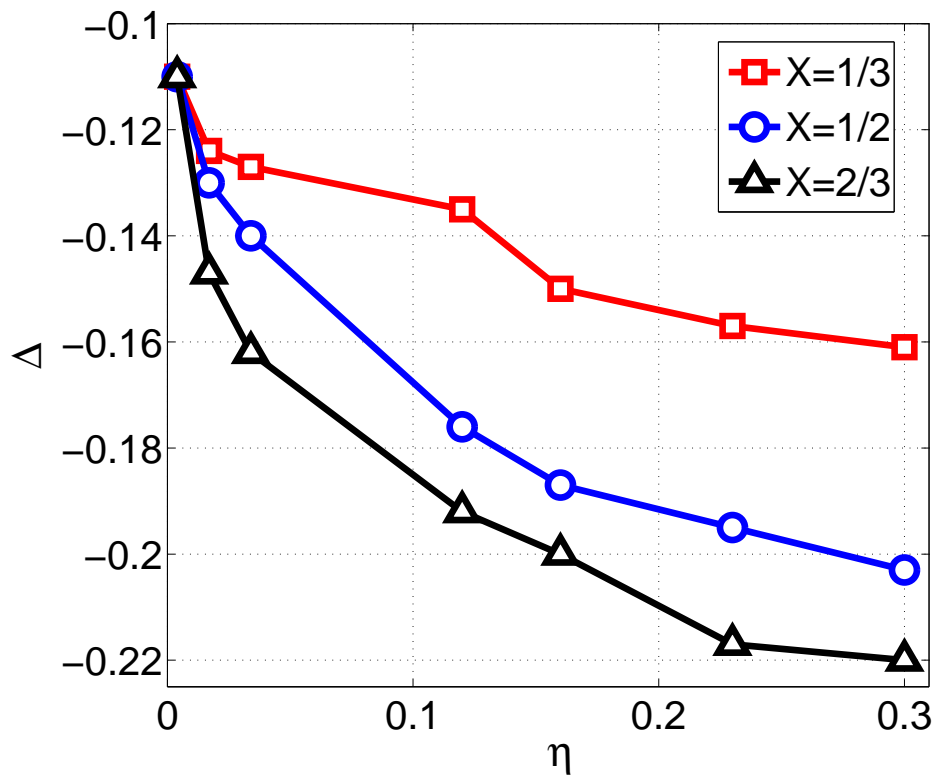}
 \caption{
(Color online) Nonadditivity parameter $\Delta$ as a function of a total macroion
   packing fraction $\eta$ for three different compositions $X=1/3, 1/2, 2/3$ . 
\label{fig-nonadd}
}
\end{figure}

\begin{figure} [!ht]
\hspace{-1.cm}
\includegraphics*[width=0.95\textwidth]{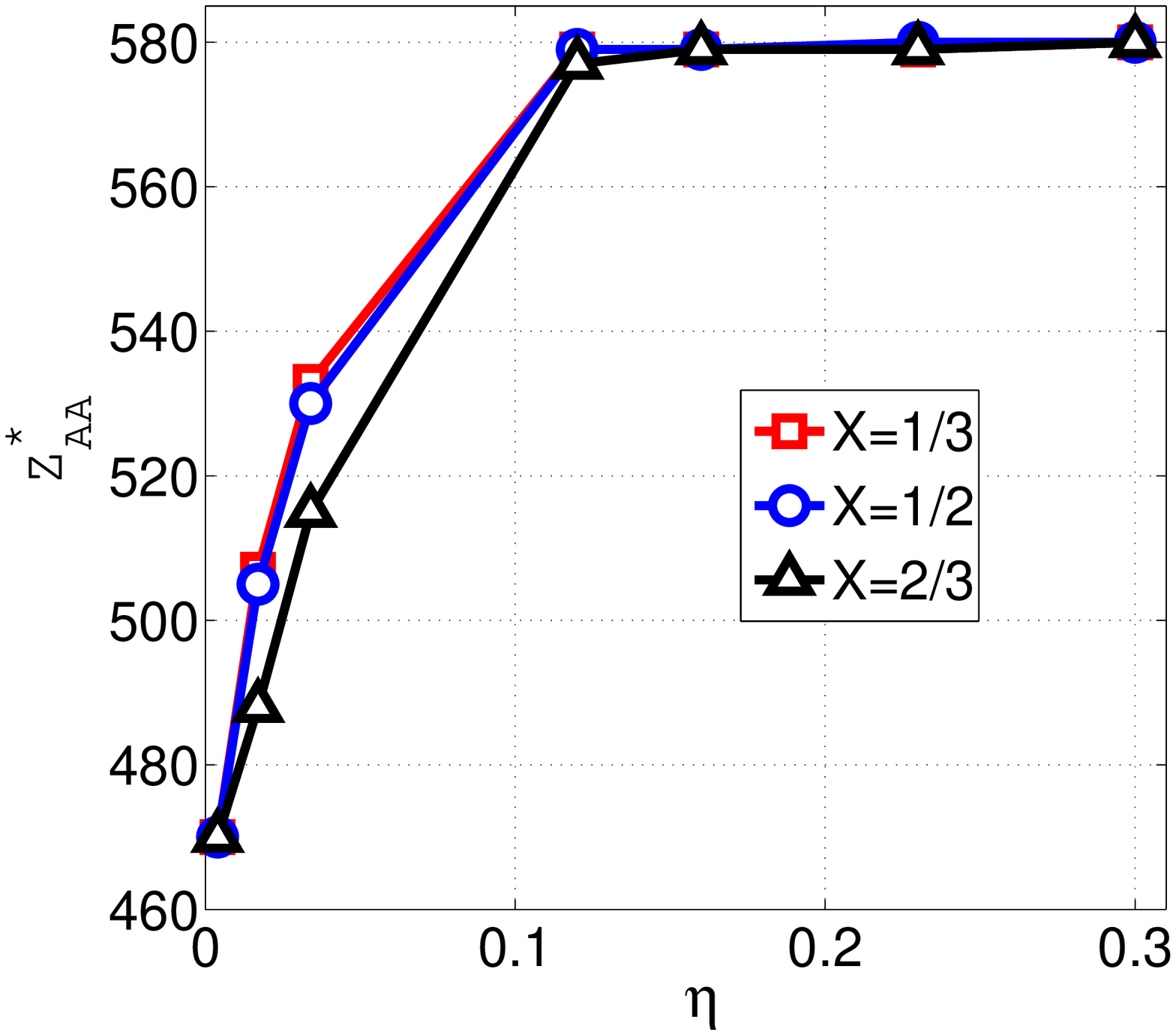}
 \caption{
(Color online) Optimal effective $AA$ charge number  $Z_{AA}^*$  as a
   function of a total macroion
   packing fraction $\eta$ for three different compositions $X=1/3, 1/2, 2/3$.
\label{fig-z1}
}
\end{figure}

\begin{figure} [!ht]
\hspace{-1.cm}
\includegraphics*[width=0.95\textwidth]{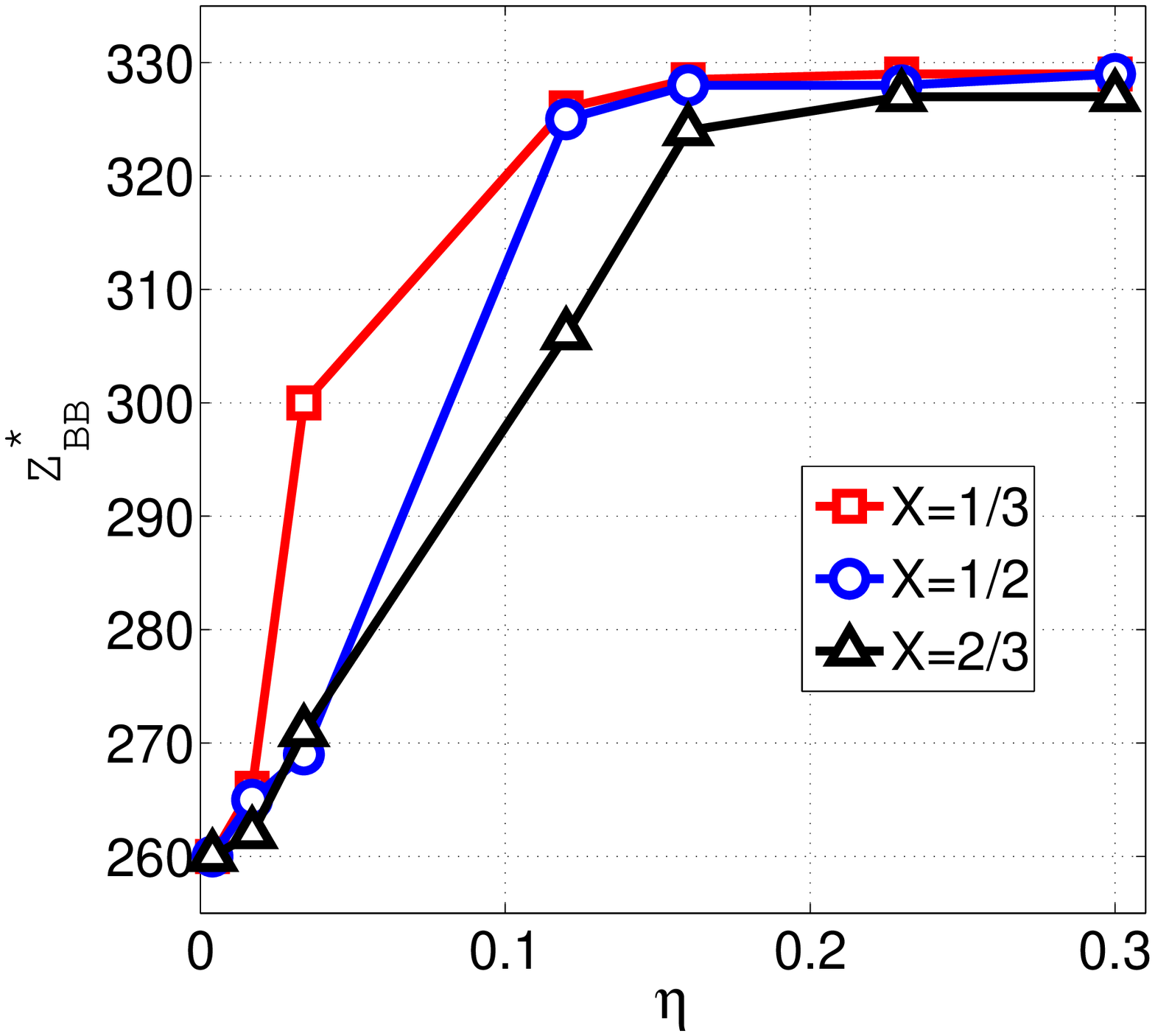}
 \caption{
(Color online) Optimal effective $BB$ charge number  $Z_{BB}^*$  as
   a function of a total macroion
   packing fraction $\eta$ for three different compositions $X=1/3, 1/2, 2/3$.
\label{fig-z2}
}
\end{figure}

\begin{figure} [!ht]
\hspace{-1.cm}
\includegraphics*[width=0.95\textwidth]{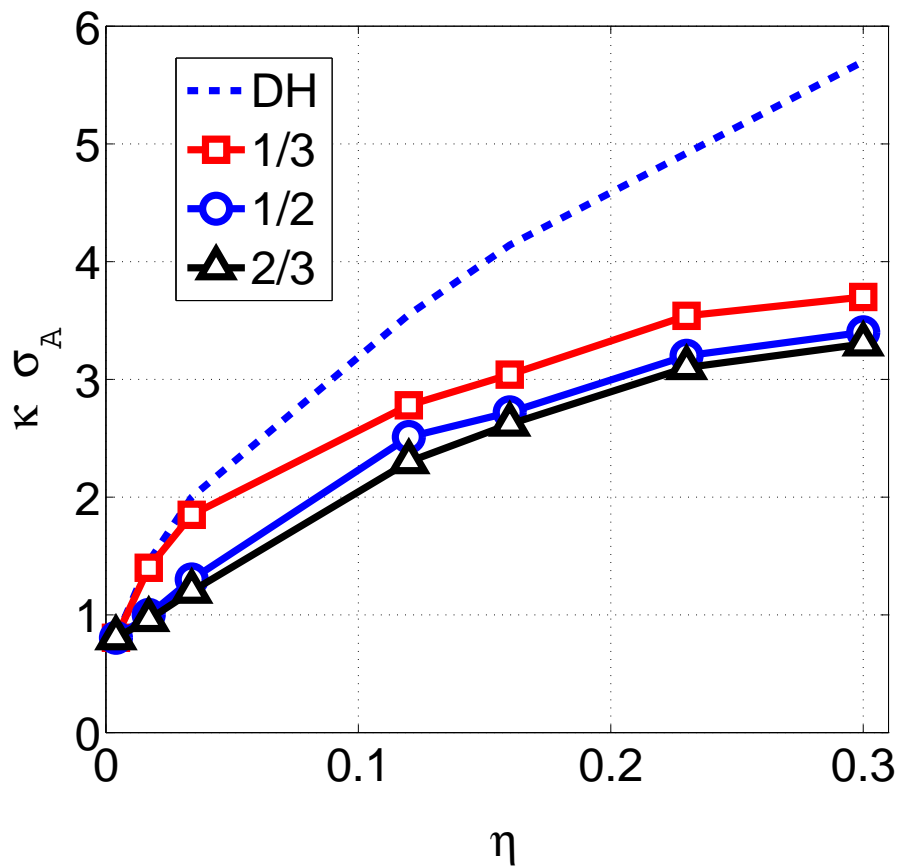}
 \caption{
(Color online) Optimal screening parameter $\kappa \sigma_A$  as
   a function of a total macroion
   packing fraction $\eta$ for three different compositions $X=1/3, 1/2, 2/3$.
The dashed line is the Debye-H\"uckel value of screening in the simulated system
according to Eq.(2) in the text. 
\label{fig-kd}
}
\end{figure}

The results of the optimal fit
for $\Delta$, ${Z_{AA}^*}$, ${Z_{BB}^*}$, and $\kappa_D \sigma_A$ are shown in 
Figures~\ref{fig-nonadd}--\ref{fig-kd} as a function of the varied macroion volume fraction $\eta$
for three different compositions $X=1/3, 1/2, 2/3$.

The nonadditivity shown in  Figure~\ref{fig-nonadd} is clearly negative and decreases with 
increasing packing fraction. This trend can be intuitively understood since asymmetries
are amplified if one approaches smaller interparticle distances. A second important 
conclusion from  Figure~\ref{fig-nonadd} is that the many-body
simulations yield the {\it same\/} value for  
$\Delta$ as those obtained from the simulations of pairs in Sect. II. 
In fact, the value $\Delta=-0.13$ is reproduced at low volume
fractions $\eta=0.017$. The effective charges and
screening length deduced from pair macroion simulations for $\eta=0.017$
perfectly fit the simulation data for many macroions at the same
packing fraction $\eta$ for the macroions. 
The effective charge numbers ${Z_{AA}^*}$ and ${Z_{BB}^*}$ shown in 
Figures~\ref{fig-z1} and \ref{fig-z2} increase slightly with volume fraction
which is the standard trend also for one-component charged suspensions
\cite{levin-2001,denton-2008}. 
The screening constant shown in Figure~\ref{fig-kd} increases with $\eta$ following the same
trend as its Debye-H\"uckel expression which is also indicated in Figure~\ref{fig-kd}.

\section{Simulations using the optimal effective Yukawa interaction}
We finally explore the impact of a non-vanishing $\Delta$ on the fluid 
structure of binary charged suspensions. In doing so we make use of the optimal
effective Yukawa fit gained in Sect. III and use it as an input in classical
coarse-grained binary Yukawa simulations (without any microions). 
These simulations can be done for much larger systems and we included 1000 $A$ and
1000 $B$ particles at equimolar composition. Two volume fraction are considered,
a dilute system with $\eta$=0.034  and
a dense system where $\eta$=0.3. 

The effective macroion charge numbers, the screening constant
and the nonadditivity parameter were chosen from results described in the previous section. 
In detail, for
the dilute system: $Z_{AA}^*$=
530,     $Z_{BB}^*$= 269, $\kappa_D\sigma_A$=1.3, $\Delta$=-0.14, 
while for the dense system, $Z_{AA}^*$=580,  $Z_{BB}^*$=330,
$\kappa_D\sigma_A$=3.4, $\Delta$=-0.2.  
At these parameters the system is in the fluid phase.

\begin{figure} [!ht]
\hspace{-1.cm}
\includegraphics*[width=0.95\textwidth]{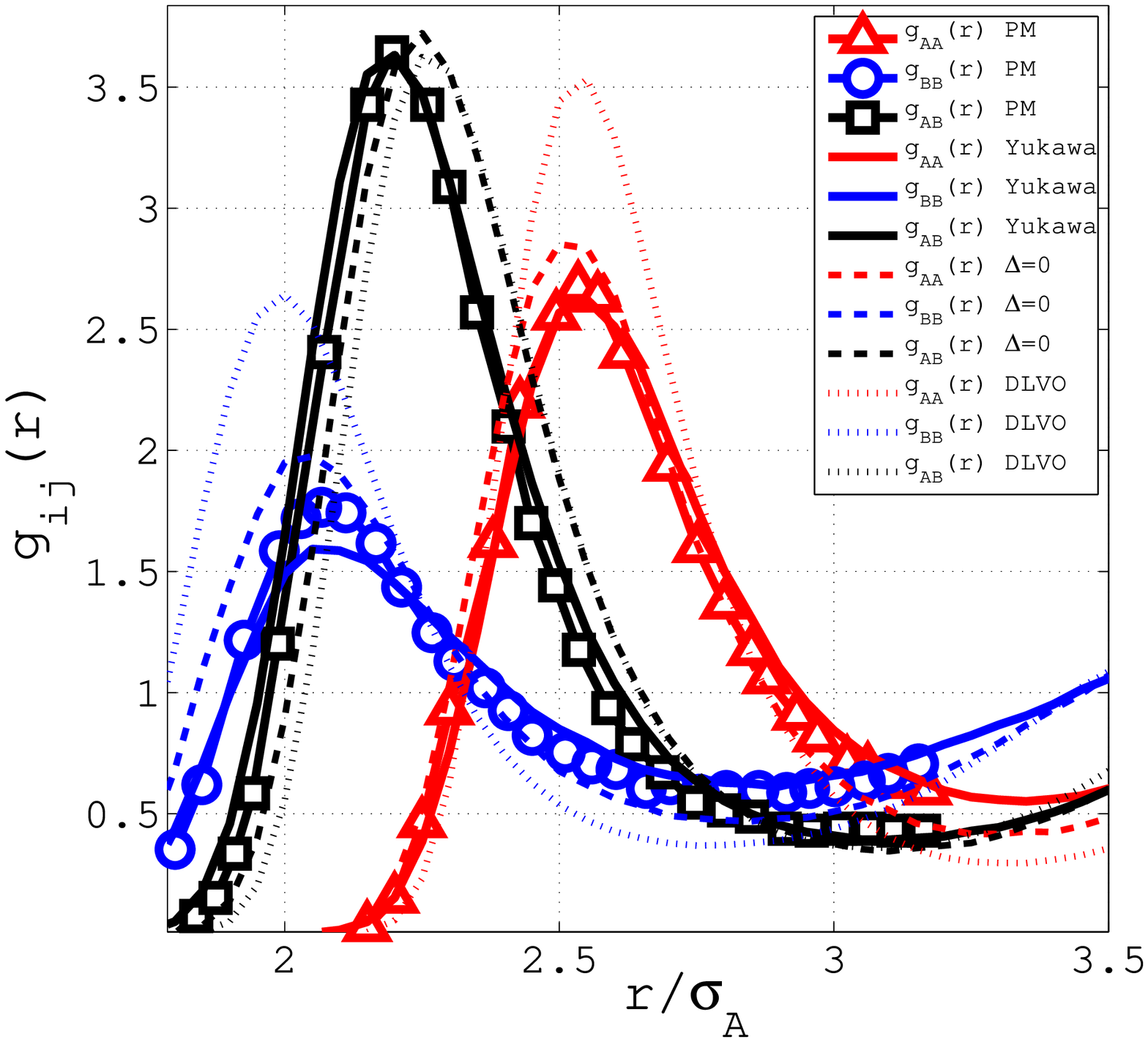}
 \caption{
(Color online) Partial macroion-macroion pair correlation functions
   for both the full primitive model  
(symbols)  and the substitute Yukawa system (full lines) for a total
packing fraction $\eta$=0.034 at composition $X=1/2$. The Yukawa simulations
   were carried for $\Delta$=-0.14, $\kappa_D \sigma_A$=1.3, $Z_{AA}^*$=530,
   $Z_{BB}^*$=269.  The additive Yukawa system results for $\Delta$=0
   and the same $\kappa_D \sigma_A$, $Z_{AA}^*$, $Z_{BB}^*$ 
   are given as dashed lines. The DLVO predictions are included as dotted lines. 
\label{fig-dilute}
}
\end{figure}

\begin{figure} [!ht]
\hspace{-1.cm}
\includegraphics*[width=0.95\textwidth]{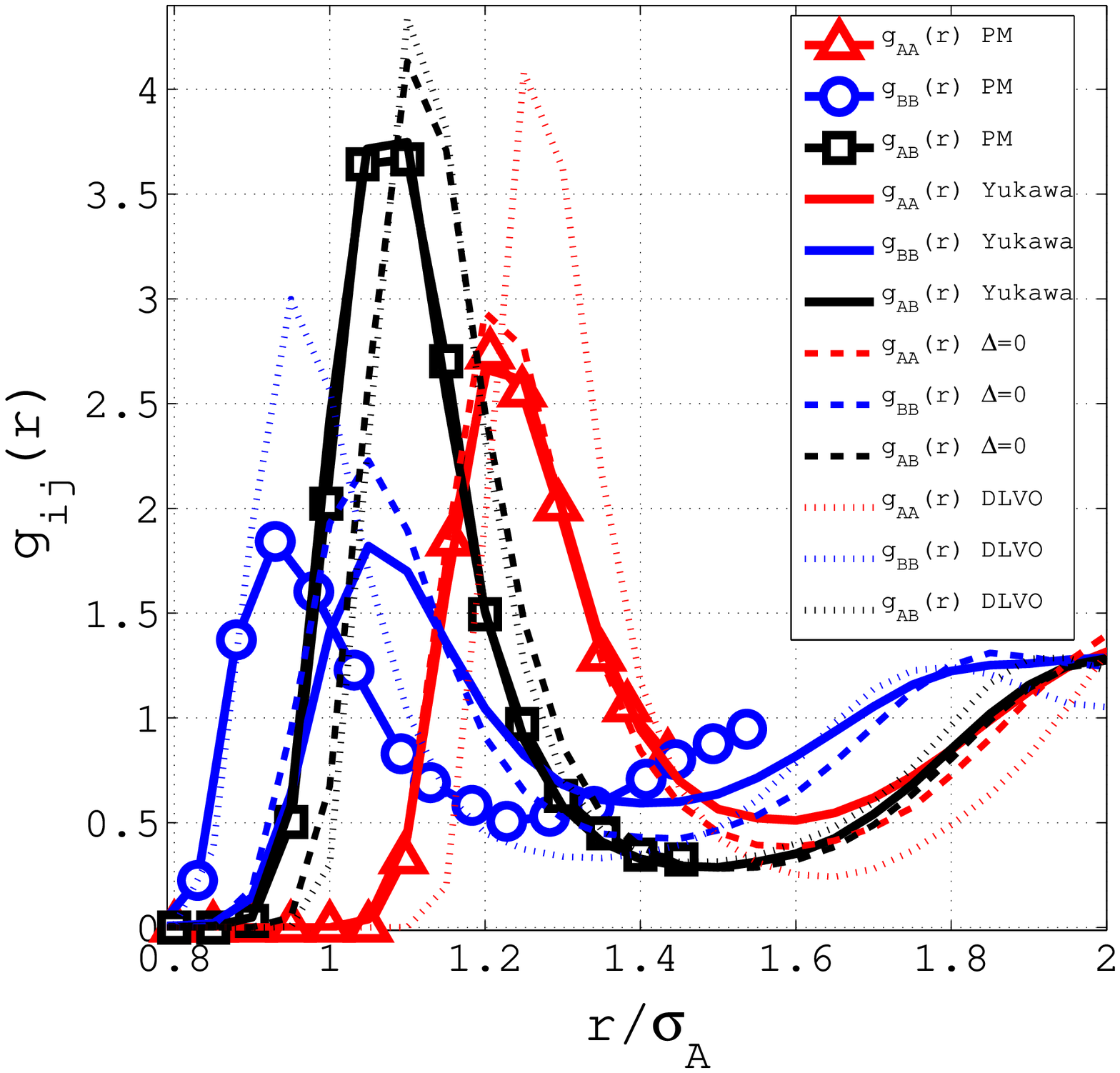}
 \caption{
(Color online)Partial macroion-macroion pair correlation functions for both the full primitive model 
(symbols)  and the substitute Yukawa system (full lines) for a total
packing fraction $\eta$=0.3 at composition $X=1/2$. The Yukawa simulations
   were carried for $\Delta$=-0.2, $\kappa_D \sigma_A$=3.4, $Z_{AA}^*$=580,
   $Z_{BB}^*$=330.  The additive Yukawa system results for $\Delta$=0
   and the same $\kappa_D \sigma_A$, $Z_{AA}^*$, $Z_{BB}^*$  are given
   as dashed lines. The DLVO predictions are included as dotted lines. 
\label{fig-dense}
}
\end{figure}

We have calculated the partial pair correlations  $g_{AA}$, $g_{BB}$ and
$g_{AB}$ for the large Yukawa substitute system and the smaller system
with explicit microions. The results are presented 
 in Figures~\ref{fig-dilute} and \ref{fig-dense}. There is good
agreement for the dilute case demonstrating that the Yukawa fit is
 reproducing the pair correlations. For the dense system, there is again agreement
except for the location of the first peak in $g_{BB}(r)$. This could have to do with 
finite-size effects of the $N=24$ primitive model systems which are expected to get 
more prominent at high packing fractions.

As a reference we have also performed Yukawa simulations with effective interactions 
based on two effective charges $Z_{AA}^*$ and $Z_{BB}^*$ but where $\Delta$ is set to zero,
i.e. where ${Z_{AB}^*}^2 = Z_{AA}^* Z_{BB}^*$. The differences in the pair structure are
pointing to the importance of nonzero nonadditivity.
There are even more deviation of the fluid pair structure when the simple DLVO expression is taken,
which significantly overestimates the structure,
see the dotted lines in Figures~\ref{fig-dilute} and \ref{fig-dense}.
 One of the main conclusions therefore is that
one has to be careful if  $\Delta$  for binary charged suspensions is neglected.

\section{Concluding remarks}
In conclusion we have determined the non-additivity parameter in binary charged
suspensions by primitive model computer simulations with explicit microions
and found significant deviations from zero. The sign
depends on the actual parameter combination, in particular on the charge asymmetry.
This  implies that a realistic 
modelling of charged suspensions
on the effective pairwise Yukawa level should incorporate a non-vanishing $\Delta$.

A priori an intuition for the sign of $\Delta$ is difficult. Intuition
only works in limiting cases. For example,
in the depletion limit of many small macroions and few big ones, there is attraction
between the big ones which would result in a positive $\Delta$.

In the future, more detailed investigation are planned to explore the full parameter space better.
It would be interesting to calculate the effective interaction in the solid phase relative to the 
fluid crystalline phase.

In inhomogeneous situations like sedimentation in a gravitational field, 
binary suspensions have been examined by primitive model computer simulations
 \cite{Ansgar} and a simple binary Yukawa pairwise interaction model was found
 to be inappropriate \cite{Roij}. Since in our bulk
simulation the pairwise Yukawa model was a good fit, we 
 expect that this is due to the density gradient in the system but this will
require more detailed studies.

Binary suspension can be prepared in a controlled way \cite{Nagele1,Stipp}
and their structural correlation, dynamics and phase diagrams  can be explored.
Depending on the size and charge ratio different freezing diagrams
 (azeotropic, spindle, eutectic) \cite{Zukoski,Meller,Wette,Lorenz} have been 
obtained in the experiments. Since the nonadditivity $\Delta$ depends on the composition,
this might point to the fact that details of the variation of $\Delta$ with 
composition and also with the phases itself (fluid or crystal) determines the shape 
of the freezing diagrams. 
Finally it would be interesting to study nonadditivity effects in charged mixtures of rods
 \cite{HL_94} or rod and sphere mixtures. It would also be interesting to
explore the effect of multivalent counterions which could lead to mutual attraction  
\cite{Amico2,DNA}.

\ack
We thank T. Palberg and A. V. Ivlev for helpful discussions.
This work was supported by the DFG via the SFB TR6 (project D1).


\newpage
\clearpage

\newpage

\newpage

\newpage

\newpage

\begin{thebibliography}{98}
\bibitem{Tammann}
G. Tammann, Ann. d. Physik. {\bf 40},  237  (1913).
\bibitem{Hafner}
J. Hafner, {\em From Hamiltonians to Phase Diagrams} (Springer, Berlin, 1987).
\bibitem{book}
G. Gompper and M. Schick, {\em Soft Matter, Vol. 2: Complex Colloidal
  Suspensions} (WILEY-VCH Verlag GmbH and Co. KGaA, Weinheim, 2006).
\bibitem{Pronk}
S. Pronk and D. Frenkel, Phys. Rev. Lett. {\bf 90},  255501  (2003).
\bibitem{Xu}
H. Xu and M. Baus, J. Phys: Condens. Matter {\bf 4},  L663  (1992).
\bibitem{Eldridge1}
M.~D. Eldridge, P.~A. Madden, and D. Frenkel, Nature {\bf 365},  35  (1993).
\bibitem{Bartlett1}
P. Bartlett, R.~H. Ottewill, and P.~N. Pusey, Phys. Rev. Lett. {\bf 68},  3801
  (1992).
\bibitem{Leunissen_Nature_2005}
M. E. Leunissen, C. G. Christova, A. P. Hynninen, C. P. Royall, A. I. Campbell, 
A. Imhof, M. Dijkstra, R. van Roij, and A. van Blaaderen, 
Nature {\bf 437}, 235 (2005).
\bibitem{Hynnien_PRL_2006}
A. P. Hynninen, C. G. Christova, R. van Roij, A. van Blaaderen, and M. Dijkstra,   
Phys. Rev. Lett.  {\bf 96}, 138308 (2006).
\bibitem{Assoud}
L. Assoud, R. Messina, and H. L{\"o}wen, Europhys. Letters {\bf 80},  48001
  (2007).
\bibitem{Likos_2007}
J. Fornleitner, F. Lo Verso, G. Kahl and C. N. Likos,
Soft Matter, {\bf 4} 480 (2008).
\bibitem{Assoud_2}  L. Assoud, R. Messina, H. L\"owen, 
J. Chem. Phys. {\bf 129}, 164511  (2008).
\bibitem{review} J.-P. Hansen, H. L{\"o}wen,
 Annual Reviews of Physical Chemistry, {\bf 51}, 209 (2000).
\bibitem{Levin} Y. Levin, Reports on Progress in Physics {\bf 65}, 1577 (2002). 
\bibitem{Messina} R. Messina, J. Phys.: Condensed Matter {\bf 21}, 113102 (2009).
\bibitem{Klein1} J. M. Mendez-Alcaraz, B. D'Aguanno, R. Klein, Physica A, {\bf 178}, 421 (1991).
\bibitem{Klein2} R. Krause, B. D'Aguanno, J. M. Mendez-Alcaraz,  R. Klein, 
J. Phys.: Condensed Matter {\bf 3}, 4459 (1991).
\bibitem{LMH2}  H.\ L{\"o}wen, J.\ P.\ Hansen, P.\ A.\ Madden,  J.\ Chem.\ Phys. {\bf 98}, 3275 
(1993).  
\bibitem{LMH1}  H.\ L{\"o}wen, P.\ A.\ Madden, J.\ P.\ Hansen, Phys.\ Rev.\ Letters {\bf 68}, 1081
 (1992).
\bibitem{triplet} H. L{\"o}wen,  E. Allahyarov, J. Phys.: Condensed Matter {\bf 10}, 4147 (1998).
\bibitem{Wu} J. Z. Wu, D. Bratko, H. W. Blanch, J. M. Prausnitz, J. Chem. Phys. {\bf 113},  3360
(2000).
J. Phys.: Condensed Matter {\bf 10}, 4147 (1998).
\bibitem{Reinke} D. Reinke, H. Stark, H. H. von Gr\"unberg, A. B. Schofield, G. Maret, U. Gasser,
Phys. Rev. Letters {\bf 98},  038301  (2007).
\bibitem{Russ} C. Russ, M. Brunner, C. Bechinger, H. H. Von Gr\"unberg,
Europhysics Letters {\bf 69},  468 (2005). 
\bibitem{Kreer} T. Kreer, J. Horbach, A. Chatterji, Phys. Rev. E {\bf 74},  021401 (2006). 
\bibitem{Trizac} E. Trizac, L. Bocquet, M. Aubouy, H. H. von Gr\"unberg, Langmuir {\bf 19}, 4027
(2003).
\bibitem{Torres} A. Torres, G. T\'ellez, R. van Roij, J. Chem. Phys. {\bf 128}, 154906 (2008).
\bibitem{nonadd_1} R. Roth, R. Evans, A. A. Louis, Phys. Rev. E {\bf 64},  051202  (2001).
\bibitem{nonadd_2} A. A. Louis, R. Roth, J. Phys.: Condensed Matter {\bf 13}, L777 (2001).
\bibitem{nonadd_3} G. Pellicane, F. Saija, C. Caccamo, P. V. Giaquinta, J. Chem. Phys. {\bf 110},  4359 (2006).

\bibitem{Hoffmann2} N. Hoffmann, F. Ebert, C.~N. Likos, G.~Maret, and H.~L{\"o}wen, Phys. Rev. Lett.
  {\bf 97},  078301  (2006).
\bibitem{Hopkins1}
P. Hopkins, A.~J. Archer, R. Evans, J. Chem. Phys. {\bf 124},  054503
  (2006).
Phys. {\bf 110},  4359 (2006).
\bibitem{Hopkins2}
P. Hopkins, A.~J. Archer,  R. Evans, J. Chem. Phys. {\bf 129},   214709  (2008). 
\bibitem{Vaulina}
O.~S. Vaulina and I.~E. Dranzhevskii, Plasma. Phys. Reports {\bf 33},  494
  (2007).
\bibitem{Kalman_PRL_2004}
G.~J. Kalman, P. Hartmann, Z. Donko, and M. Rosenberg, Phys. Rev. Lett. {\bf
  92},  065001  (2004).
\bibitem{Sutterlin} K. R. S\"utterlin, A. Wysocki, A. V. Ivlev, C. R\"ath, H. M. Thomas,  M. Rubin-Zuzic, 
 W. J. Goedheer, V. E. Fortov, A. M. Lipaev, V. I. Molotkov, O. F. Petrov, 
G. E. Morfill, H. L\"owen,
Phys. Rev. Letters {\bf 102}, 085003  (2009).
\bibitem{Horbach1} N. Kikuchi, J. Horbach, EPL {\bf 77},  26001  (2007).
\bibitem{Horbach2} A. Carre, L. Berthier, J. Horbach, S. Ispas, W. Kob,
J. Chem. Phys. {\bf 127},  114512 (2007).
\bibitem{Louis}
A.~A. Louis, E. Allahyarov, H. L{\"o}wen, and R. Roth, Phys. Rev. E {\bf 65},
  061407  (2002).
\bibitem{phase1}
E. Sch\"oll-Paschinger and G. Kahl, J. Chem. Phys. {\bf 118},  7414  (2003).
\bibitem{phase2}
P. Hopkins, A.~J. Archer, and R. Evans, J. Chem. Phys. {\bf 124},  054503
  (2006).
\bibitem{phase3}
J. K\"ofinger, N.~B. Wilding, and G. Kahl, J. Chem. Phys. {\bf 125},  234503
  (2006).
\bibitem{polydis}  H.\ L{\"o}wen, J.\ P.\ Hansen, J.\ N.\ Roux,  Phys.\ Rev.\ A {\bf 44},
1169 (1991).
\bibitem{dynamics1}
M.~A. Chavez-Rojo and M. Medina-Noyola, Physica A {\bf 366},  55  (2006).
\bibitem{dynamics2}
N. Kikuchi and J. Horbach, Europhys. Letters {\bf 77},  26001  (2007).
\bibitem{Chavez} M.~A. Chavez-Rojo, R. Juarez-Maldonado,  
M. Medina-Noyola, Phys. Rev. E {\bf 77}, 040401(R) (2008).
\bibitem{Medina} R. Juarez-Maldonado, M. Medina-Noyola, Phys. Rev. E {\bf 77}, 051503 (2008). 
\bibitem{Salin}
G. Salin and D. Gilles, J. Phys. A: Math. Gen. {\bf 17},  4517  (2006).
\bibitem{Ivlev} A. V. Ivlev, S. K. Zhdanov, H. M. Thomas, G. E. Morfill,  EPL (in press).
\bibitem{Amico1} I. D'Amico, H. L{\"o}wen,  
 Physica A {\bf 237}, 25 (1997). 
\bibitem{Amico2} E. Allahyarov, I. D'Amico, H. L{\"o}wen,
 Phys. Rev. Letters {\bf 81}, 1334 (1998).
\bibitem{Trigger}
E. Allahyarov, H. L{\"o}wen, and S. Trigger, Phys. Rev. E {\bf 57},  5818
  (1998).
\bibitem{Damico}
E. Allahyarov, I. D'Amico, and H. L{\"o}wen, Phys. Rev. E {\bf 90}, 3199 (1999).
\bibitem{Kramposthuber} H.\ L{\"o}wen, G.\ Kramposthuber,   Europhys. Letters {\bf 23}, 637
 (1993).
\bibitem{wedge} H. L{\"o}wen, A. H\"artel, A. Barreira-Fontecha, 
H. J. Sch\"ope, E. Allahyarov, T. Palberg, 
 J. Phys.: Condensed Matter  {\bf 20}, 404221  (2008).
\bibitem{protein1} E. Allahyarov, H. L{\"o}wen, A. A.  Louis, J.-P. Hansen,
 Europhys. Letters {\bf 57}, 731 (2002).
\bibitem{protein2} E. Allahyarov, H. L{\"o}wen, A. A.  Louis, J.-P. Hansen,
Phys. Rev. E {\bf 67},  051404  (2003)
\bibitem{Zacca} E. Allahyarov, E. Zaccarelli, F. Sciortino, P. Tartaglia,
H. L{\"o}wen,
Europhysics Letters {\bf 78}, 38002  (2007).
\bibitem{levin-2001} A. Diehl, M. C. Barbosa, Y. Levin, Europhysics
  Letters {\bf 53}, 86 (2001).
\bibitem{denton-2008} A. R. Denton, J. Phys.: Cond. Matter {\bf 20},
  494230 (2008).
\bibitem{Ansgar}  A. Esztermann, H. L{\"o}wen,
Europhys. Letters {\bf 68}, 120 (2004).
\bibitem{Roij} A. Torres, A. Cuetos, M. Dijkstra, R. van Roij, Phys. Rev. E {\bf 77}, 
 031402  (2008).
\bibitem{Nagele1} G. H. Koenderink, H. Y. Zhang, M. P. Lettinga, G. N\"agele, A. P. Philipse,
Phys. Rev. E {\bf 64},  022401  (2001).
\bibitem{Stipp} A. Stipp, T. Palberg, Phil. Mag. Letters {\bf 87}, 899 (2007).
\bibitem{Zukoski} W. J. Hunt, C. F. Zukoski, J. Colloid Interface Sci. {\bf 210}, 332 (1999).
\bibitem{Meller} A. Meller, J. Stavans, Phys. Rev. Letters {\bf 68}, 3645 (1992)
\bibitem{Wette} P. Wette, H. J. Sch\"ope, T. Palberg, J. Chem. Phys. {\bf 122},  144901 (2005).
\bibitem{Lorenz} N. Lorenz, J. N. Liu, T. Palberg, Colloids and Surfaces A {\bf 319}, 109 (2008).
\bibitem{HL_94}  H.\ L{\"o}wen,  J. Chem. Phys. {\bf 100}, 6738 (1994).
\bibitem{DNA}
 E. Allahyarov, G. Gompper, H. L{\"o}wen,
  Phys. Rev. E {\bf 69}, 041904  (2004).

\end{thebibliography}
\end{document}